\documentclass[letterpaper]{article} 
\usepackage{aaai24}  
\usepackage{times}  
\usepackage{helvet}  
\usepackage{courier}  
\usepackage[hyphens]{url}  
\usepackage{graphicx} 
\usepackage{natbib}  
\usepackage{caption} 
\usepackage{newfloat}
\usepackage{listings}

\usepackage{xcolor}
\usepackage{amsmath}
\usepackage{amssymb}
\usepackage{bibentry}
\usepackage[inline]{enumitem}
\usepackage{algorithm}
\usepackage{algorithmic}
\usepackage{multirow}
\usepackage{booktabs}

\newtheorem{theorem}{Theorem}

\newtheorem{definition}{Definition}[section]

\DeclareCaptionStyle{ruled}{labelfont=normalfont,labelsep=colon,strut=off} 
\floatstyle{ruled}
\newfloat{listing}{tb}{lst}{}
\floatname{listing}{Listing}
\newcommand{\heading}[1]{\vspace*{0.8mm}\noindent\textbf{#1.}}

\title{Perturbation-Invariant Adversarial Training for Neural Ranking Models: \\ Improving the Effectiveness-Robustness Trade-Off}
\author {
    Yu-An Liu\textsuperscript{\rm 1,\rm 2},
    Ruqing Zhang\textsuperscript{\rm 1,\rm 2},
    Mingkun Zhang\textsuperscript{\rm 1,\rm 2},
    Wei Chen\textsuperscript{\rm 1,\rm 2},
    Maarten de Rijke\textsuperscript{\rm 3}, \\
    Jiafeng Guo\textsuperscript{\rm 1,\rm 2},
    Xueqi Cheng\textsuperscript{\rm 1,\rm 2}
}
\affiliations {
    \textsuperscript{\rm 1}CAS Key Lab of Network Data Science and Technology, Institute of Computing Technology, \\Chinese Academy of Sciences, Beijing, China\\
    \textsuperscript{\rm 2}University of Chinese Academy of Sciences, Beijing, China\\
    \textsuperscript{\rm 3}University of Amsterdam, Amsterdam, The Netherlands\\
    \{liuyuan21b, zhangruqing, zhangmingkun20z, chenwei2022, guojiafeng, cxq\}@ict.ac.cn, m.derijke@uva.nl 
}

\begin{document}

\maketitle

\begin{abstract}
Neural ranking models (NRMs) have shown great success in information retrieval (IR).
But their predictions can easily be manipulated using adversarial examples, which are crafted by adding imperceptible perturbations to legitimate documents. 
This vulnerability raises significant concerns about their reliability and hinders the widespread deployment of NRMs.
By incorporating adversarial examples into training data, adversarial training has become the de facto defense approach to adversarial attacks against NRMs. 
However, this defense mechanism is subject to a trade-off between effectiveness and adversarial robustness. 
In this study, we establish theoretical guarantees regarding the effectiveness-robustness trade-off in NRMs. 
We decompose the robust ranking error into two components, i.e., a \emph{natural ranking error} for effectiveness evaluation and a \emph{boundary ranking error} for assessing adversarial robustness. 
Then, we define the \emph{perturbation invariance} of a ranking model and prove it to be a differentiable upper bound on the boundary ranking error for attainable computation. 
Informed by our theoretical analysis, we design a novel \emph{perturbation-invariant adversarial training} (PIAT) method for ranking models to achieve a better effectiveness-robustness trade-off. 
We design a regularized surrogate loss, in which one term encourages the effectiveness to be maximized while the regularization term encourages the output to be smooth, so as to improve adversarial robustness. 
Experimental results on several ranking models demonstrate the superiority of PITA compared to existing adversarial defenses. 
\end{abstract}


\section{Introduction}

Ranking is a fundamental problem in information retrieval (IR). 
With advances in deep learning \cite{lecun2015deep}, neural ranking models (NRMs) \citep{guo2020deep} have achieved remarkable effectiveness.
We have also witnessed substantial uptake of NRMs in practice~\cite{lin2022pretrained}.
Recently, it has been demonstrated that NRMs are vulnerable to adversarial examples that are capable of inducing misbehavior with human-imperceptible perturbations \cite{wu2022prada,liu2022order,chen2023towards}.
So far, little attention has been devoted to combating this issue. 
A representative and successful method for attacking NRMs is the word substitution ranking attack (WSRA), which promotes a target document in rankings by replacing important words with synonyms \cite{wu2022prada}.
Given the prevalence of black-hat search engine optimization (SEO)~\cite{gyongyi2005web}, enhancing the adversarial robustness of NRMs against such attacks is vital for their use in real-world scenarios. 

Among adversarial defense mechanisms proposed to improve model robustness  \cite{jia2017adversarial,raghunathan2018certified,madry2018towards}, \emph{adversarial training} remains the top-performer \cite{shafahi2019adversarial,zhu2019freelb}. 
During adversarial training adversarial examples are fed to a model. 
However, this causes an undesirable reduction in effectiveness on natural (clean) samples, giving rise to a trade-off dilemma between effectiveness and robustness \cite{tsipras2019robustness}. 
This is because effectiveness concerns the overall performance under normal conditions, while adversarial robustness centers on performance under malicious behavior.
Several refinements have been suggested for vanilla adversarial training, to mitigate the aforementioned trade-off in text and image classification  \cite{zhang2019theoretically, wang2021adversarial}. 
However, clear differences exist between classification and ranking scenarios concerning the trade-off, given that the former relies on a single sample, whereas the latter involves a ranked list. 
So far, the ranking task has not benefited from these advances in bridging the gap between effectiveness and robustness. 
This naturally raises the first question:
\begin{quote}
    \textit{What is the trade-off between effectiveness and robustness for ranking problems?}
\end{quote}
We contribute a theoretical characterization of this question by decomposing the robust ranking error, i.e., the prediction error for adversarial examples, into two terms: 
\begin{enumerate*}[label=(\roman*)]
    \item a \emph{natural ranking error}, which focuses on the natural effectiveness of the ranked list predicted by the ranking model on clean data, and 
    \item a \emph{boundary ranking error}, which indicates the ranking model's adversarial robustness against adversarial examples, measuring the proximity of input features to the decision boundary. 
\end{enumerate*}
We then introduce the \emph{perturbation invariance} of a ranking model, which says that any adversarial perturbation to candidate documents does not alter the resulting document ranking. 
We prove that the perturbation invariance is a differentiable upper bound on the boundary ranking error, which is sufficiently tight. 
Differences in measurements of these two errors, which express distinct optimization objectives, showcase the trade-off between effectiveness and robustness for ranking problems.

Next to the effectiveness-robustness trade-off, the second issue we address is:
\begin{quote}
    \textit{How to design a defense mechanism against adversarial examples while maintaining competitive effectiveness for NRMs guided by our theoretical characterization?}
\end{quote}
We introduce a novel \emph{perturbation-invariant adversarial training method} (PIAT) to achieve this goal. 
The key idea is to capture the trade-off between natural and boundary ranking error by optimizing a regularized surrogate loss composed of two terms: 
\begin{enumerate*}[label=(\roman*)]
    \item a \emph{natural ranking loss}, which encourages the optimization of the natural ranking error by minimizing the ``difference'' between the predicted ranked list and the ground-truth based on supervised data, and 
    \item an \emph{adversarial ranking loss}, as the regularization term, which encourages the optimization of boundary ranking error by minimizing the ``difference'' between the predicted ranked list on natural candidates and on attacked candidates using semi-supervised learning. 
\end{enumerate*}
We propose three ways to implement the regularization term to ensure perturbation invariance. 
By combining supervised and semi-supervised training, we effectively leverage information from large-scale volumes of unlabeled documents to improve the effectiveness-robustness trade-off for NRMs. 

Extensive experiments conducted on the widely-used MS MARCO passage ranking dataset show that PIAT offers superior defense against WSRA while maintaining effectiveness as compared to several empirical defense methods, including data augmentation and vanilla adversarial training. 
Ablations and visualizations are provided for more insights.

\section{Preliminaries}

Our work focuses on adversarial robustness to word substitution ranking attacks for NRMs. We review this type of attack in this section.  

\heading{Attacks in web search} 
The web, as a canonical example of a competitive search setting, involves document authors who have incentives to optimize their content for better rankings in search results \cite{kurland2022competitive}. 
This practice is commonly known as search engine optimization (SEO), which aims at improving the visibility and ranking of a web page in retrieved results when specific queries are entered by users \cite{gyongyi2005web}. 
This can lead to a decrease in the overall quality of search results, as many irrelevant or low-quality documents may end up being ranked higher than they deserve, while more valuable and accurate content may get pushed down in the results.

\heading{Word substitution ranking attack}
Recently, there has been much research on adversarial attacks against NRMs to simulate real-world ranking competitions. 
A representative study is the word substitution ranking attack (WSRA) \cite{wu2022prada}, which  demonstrates promising results in terms of the attack success rate. 
Given a ranking model, WSRA aims to promote a target document in rankings by replacing important words in its text with synonyms in a semantics-preserving way. 
Our research concentrates on WSRA attacks and develops a corresponding defense strategy. 

Typically, in ad-hoc retrieval, given a query $\boldsymbol{q}$ and a set of document candidates  $\mathcal{D} = \left\{ \boldsymbol{d}_{1}, \boldsymbol{d}_{2}, \ldots, \boldsymbol{d}_{N_d} \right\}$, a neural ranking model $f$ predicts the relevance score $f\left(\boldsymbol{q},\boldsymbol{d}_i\right)$ of each query-document pair for ranking the whole candidate set.
For example, $f$ outputs the ranked list $[\boldsymbol{d}_{N_d}, \boldsymbol{d}_{N_d-1},\ldots, \boldsymbol{d}_1]$ if it determines $f\left(\boldsymbol{q},\boldsymbol{d}_{N_d}\right) > f\left(\boldsymbol{q},\boldsymbol{d}_{N_d-1}\right) > \cdots > f\left(\boldsymbol{q},\boldsymbol{d}_1\right)$.
The rank position of document $\boldsymbol{d}_{i}$ with respect to query $\boldsymbol{q}$ predicted by $f$ is $\pi_{f}\left(\boldsymbol{q}, \boldsymbol{d}_{i}\right)$.
And we use $\pi_{y}\left(\boldsymbol{q}, \boldsymbol{d}_{i}\right)$ to represent the ground-truth rank position of $\boldsymbol{d}_{i}$ with respect to $\boldsymbol{q}$.

Given a target document $\boldsymbol{d}= \left(w_1, w_2, \ldots, w_M \right) \in \mathcal{D}$,  the WSRA task constructs an adversarial example $\boldsymbol{d}^{\prime} = \left( w_1^{\prime}, w_2^{\prime}, \ldots, w_M^{\prime}\right)$ by replacing at most $\epsilon \cdot M \left(\epsilon \leq 1\right)$ words in $\boldsymbol{d}$ with any of their synonyms $w_m'$. 
We denote a candidate set of adversarial examples (neighborhood) of $\boldsymbol{d}$ as $\mathbb{B}(\boldsymbol{d}, \epsilon)$, i.e.,
\begin{equation}
\begin{aligned}
\mathbb{B}(\boldsymbol{d}, \epsilon) :=\left\{
\boldsymbol{d}^{\prime}:\left\|\boldsymbol{d}^{\prime}-\boldsymbol{d}\right\|_0 /\|\boldsymbol{d}\| \leq \epsilon \right\}
\end{aligned},
\end{equation}
where  $\left\|\boldsymbol{d}\right\|$ represents the number of words in document $\boldsymbol{d}$, $\left\|\boldsymbol{d}^{\prime} - \boldsymbol{d}\right\|_0 := \sum_{m=1}^{M} \mathbb{I}\left\{w_m' \neq w_m \right\}$ is the Hamming distance, with $\mathbb{I}\{\cdot \}$ the indicator function.
Ideally, the goal of the attacker is to find $\boldsymbol{d}^{\prime} \in \mathbb{B}(\boldsymbol{d}, \epsilon)$ such that $f(\boldsymbol{q},\boldsymbol{d}^{\prime})>f(\boldsymbol{q},\boldsymbol{d}) $ and  $\boldsymbol{d}^{\prime}$ has the same semantic meaning as $\boldsymbol{d}$. 

\section{Theoretical Analysis: The Trade-off between Effectiveness and Robustness}

\citet{tsipras2019robustness} have shown that  the goals of standard performance and adversarial robustness may be at odds. 
There can be an inherent trade-off between effectiveness and robustness.  
Drawing inspiration from the definitions of natural and robust accuracy in  \cite{zhang2019theoretically}, we characterize the trade-off in ranking by breaking down the robust ranking error into the sum of the natural ranking error and boundary ranking error. 
We also provide a differentiable upper bound on the boundary ranking error, to inform the design of the defense method.

\subsection{Natural ranking error}

So far, much effort in the field of NRMs has been dedicated to improving the ranking  effectiveness, which is about the average performance under normal conditions. 

\begin{definition}[Natural ranking error]
\rm
Formally, the \emph{natural error} associated with the effectiveness of a ranking model $f$ on natural (clean) examples is denoted as,    
\begin{equation}
    \begin{aligned}
        \mathcal{R}_{\mathrm{nat}}(f):=\mathbb{E}_{\boldsymbol{d}_{i} \sim \mathcal{D}} \mathbb{I}\left\{\pi_{f}(\boldsymbol{q}, \boldsymbol{d}_{i}) \neq  \pi_{y}(\boldsymbol{q}, \boldsymbol{d}_{i}) \right\}
    \end{aligned},
\end{equation}
where $\mathbb{I}\{ \cdot \}$ is the indicator function that is $1$ if an event happens and $0$ otherwise. 
For simplicity, we consider the $0-1$ loss in our theoretical analysis to evaluate the natural error. 
\end{definition}

\subsection{Boundary ranking error}

Here, we first define the decision boundary of a ranking model, and then introduce the boundary ranking error corresponding to the adversarial robustness of ranking models. 

\begin{definition}[Ranking decision boundary]
\rm
For a ranking model $f$, we define the \emph{ranking decision boundary} as the predicted rank position  $\pi_{f}(\boldsymbol{q}, \boldsymbol{d}_{i})$ being higher or lower than it truly deserves. 
Note that for the topmost and the bottom-most ranks, we exclusively consider the situations where the predicted rank is one position lower and higher, respectively.
Considering practical attacks aimed at ranking improvement, we denote $ \pi_{n}(\boldsymbol{q}, \boldsymbol{d}_{i}) = \pi_{y}(\boldsymbol{q}, \boldsymbol{d}_{i}) - 1$ as the neighborhood rank of $(\boldsymbol{q}, \boldsymbol{d}_{i})$. 
Recall that low values of rank positions attest to high ranking. 
In this way, the ranking decision boundary can be formulated as:
\begin{equation}
    \begin{aligned}
        \text{DB}(f) := 
        \left\{\boldsymbol{d}_{i} \sim \mathcal{D} : \pi_{f}(\boldsymbol{q}, \boldsymbol{d}_{i}) = \pi_{n}(\boldsymbol{q}, \boldsymbol{d}_{i}) \right\}.
    \end{aligned}
\end{equation}
\end{definition}
We use $\mathbb{B}(\boldsymbol{d}_{i}, \epsilon)$ to represent a neighborhood of $\boldsymbol{d}_{i}$ under the WSRA attack. 
Then, for a ranking model $f$, we denote the neighborhood of the decision boundary of $f$ as:
\begin{equation}
    \label{epsilon ball of boundary}
    \begin{split}
        & \mathbb{B}(\text{DB}(f), \epsilon) := \{ \boldsymbol{d}_{i} \sim \mathcal{D} : \ \exists ~\boldsymbol{d}_{i}^{\prime} \in \mathbb{B}(\boldsymbol{d_{i}}, \epsilon) \text{ such that} \\
        & [ \pi_{f}(\boldsymbol{q}, \boldsymbol{d}_{i}) \!-\! \pi_{n}(\boldsymbol{q}, \boldsymbol{d}_{i})   ] \cdot [ \pi_{f}(\boldsymbol{q}, \boldsymbol{d}^{\prime}_{i}) \!-\! \pi_{n}(\boldsymbol{q}, \boldsymbol{d}_{i})] \!\leq\! 0  \}.
    \end{split}
\end{equation}
This implies that $\boldsymbol{d}_{i}$ and $\boldsymbol{d}^{\prime}_{i}$ are located on different sides of the decision boundary concerning the query $\boldsymbol{q}$. 
Therefore, a successful adversarial attack could move the target document to the wrong side of the decision boundary, leading to weak robustness of NRMs. 

The above analysis elucidates why a ranking model with high effectiveness might still manifest considerable adversarial vulnerability. 
This discrepancy arises from the distinction between optimizing based on natural ranking error and acquiring a robust decision boundary for NRMs. 
Based on experimental findings due to \citet{wu2022prada}, we can tell:
\begin{enumerate*}[label=(\roman*)]
    \item decision boundaries learned based on natural ranking errors enable NRMs to achieve high effectiveness on clean documents, and 
    \item such boundaries are susceptible to being breached by adversarial examples, resulting in vulnerabilities to easy attacks. 
\end{enumerate*}
Existing attack methods take advantage of this boundary vulnerability to deceive the NRM. 
As such, training robust NRMs requires a defining boundary ranking error to tackle this vulnerability effectively.

\begin{definition}[Boundary ranking error] 
\label{def: boundary ranking error}
\rm
We introduce the \emph{boundary ranking error} to assess the existence of adversarial examples lying around the ranking decision boundary of $f$, i.e.,
\begin{equation} 
    \label{R_bdy definition}
    \begin{aligned}
        & \mathcal{R}_{\mathrm{bdy}}(f):= \\
        & \mathbb{E}_{\boldsymbol{d}_{i} \sim \mathcal{D}} \mathbb{I}\left\{ \boldsymbol{d}_{i} \in \mathbb{B} \! \left(\text{DB}(f), \epsilon\right) , \pi_{f}\!\left(\boldsymbol{q}, \boldsymbol{d}_{i}\right) = \pi_{y}\!\left(\boldsymbol{q}, \boldsymbol{d}_{i}\right) \right\}.
    \end{aligned}
\end{equation}
\end{definition}
Optimizing the boundary ranking error poses a challenge, mainly due to the large volume of unlabeled documents in the datasets and the unavailability of ground-truth rankings. 
To address this obstacle, we present a solution in the form of an upper bound on the boundary ranking error.

\begin{theorem}[Upper bound of boundary ranking error] 
\label{thm:1}
According to Eq. \ref{epsilon ball of boundary} and~\ref{R_bdy definition}, for a ranking model $f : \boldsymbol{q} \times \mathcal{D} \rightarrow \mathbb{R}$ and ranking mechanism $r : \mathbb{R} \times \mathbb{R} \rightarrow \left\{\pm 1 , 0 \right\}$, we have:
\begin{equation}
    \begin{split}
        \mathcal{R}_{\mathrm{bdy}}(f) \!
        \leq \!\mathbb{E}_{\boldsymbol{d}_{i} \sim \mathcal{D}} \!\max _{\boldsymbol{d}_{i}^{\prime} \in \mathbb{B}(\boldsymbol{d}_{i}, \epsilon)} \!\mathbb{I}\left\{ \pi_{f}(\boldsymbol{q}, \boldsymbol{d}_{i})  \neq \pi_{f}(\boldsymbol{q}, \boldsymbol{d}^{\prime}_{i}) \right\}\!.
        \hspace*{-2mm}\mbox{}
    \end{split}
\end{equation}
\end{theorem}
According to Theorem~\ref{thm:1}, the boundary ranking error can be upper-bounded by the expectation that any adversarial example is ranked in its original ranking positions.
This emphasizes the perturbation invariance of a robust ranking model, that is, any perturbation to the inputted candidate documents does not change the output ranking. 
Consequently, restraining the boundary ranking error is attainable by maximizing the outputted perturbation invariance of ranking models.

The proof for Theorem~\ref{thm:1} is presented in Appendix \ref{proof:1}. 
Nonetheless, if an upper bound is too loose, it may lead to the inadequacy of effectively optimizing the error. 
Hence, we further prove the tightness of the upper bound in Theorem~\ref{thm:1}, which is shown in Appendix \ref{proof:2}.
The procedure ensures the reduction of the boundary ranking error through the optimization of perturbation invariance.

\subsection{Trade-off between natural and boundary ranking errors}

Based on the definitions of natural error and boundary error for a ranking model, we present the robust ranking error for adversarial examples. 

\begin{definition}[Robust ranking error]
\label{Robust Ranking Error}
\rm
To train a robust ranking model, the \emph{robust ranking error $\mathcal{R}_{\mathrm{rob}}(f)$} under the WSRA scenario, can be decomposed as follows,  
\begin{equation}\label{EQ_rob_decomposition}
\begin{aligned}
\mathcal{R}_{\mathrm{rob}}(f) = \mathcal{R}_{\mathrm{nat}}(f) + \mathcal{R}_{\mathrm{bdy}}(f),
\end{aligned}
\end{equation}
where $\mathcal{R}_{\mathrm{nat}}(f)$ corresponds to naturally wrongly ranked documents; and $\mathcal{R}_{\mathrm{bdy}}(f)$ corresponds to correctly ranked samples but close to the $\epsilon$-extension of the ranking decision boundary. Consequently, these samples are susceptible to successful boundary-crossing attacks (i.e., ranked higher or lower) by introducing human-imperceptible perturbations (e.g., word substitution). 
\end{definition}

\section{Algorithmic Design: Perturbation-\\invariant Adversarial Training}

Inspired by our theoretical analysis, we present a new defense method for NRMs, named perturbation-invariant adversarial training (PIAT), to strike a balance between effectiveness and adversarial robustness.  

\subsection{Motivation}

Theorem~\ref{thm:1} and Definition~\ref{Robust Ranking Error} emphasize the importance of simultaneously optimizing the natural ranking error and the boundary ranking error, when training a robust ranking model while preserving effectiveness. 
We introduce a refinement to adversarial training, called PIAT, tailored specifically for ranking problems. 
This involves the incorporation of a \emph{regularized surrogate loss} aimed at optimizing the robust ranking error, comprising two essential terms, i.e.,  
\begin{equation}
    \label{overall-loss}
    \begin{aligned}
        \mathcal{L} = \lambda \mathcal{L}_{\mathrm{nat}} + (1 - \lambda) \mathcal{L}_{\mathrm{adv}},
    \end{aligned}
\end{equation}
where the first term, i.e., the natural ranking loss $\mathcal{L}_{\mathrm{nat}}$, encourages the natural ranking error to be optimized, by minimizing the ``difference'' between the predicted and ground-truth ranked list. 
We achieve this by leveraging a traditional pair-wise loss, which is supervised using the labeled query-document pairs. 
The regularization term, i.e., the adversarial ranking loss $\mathcal{L}_{\mathrm{adv}}$, encourages the boundary ranking error to be optimized. 
We propose a perturbation-invariant ranking loss to minimize the  ``difference'' between the prediction of a clean document set and that of an attacked document set, which is semi-supervised by the NRM's outputs. 
The $\lambda$ is a trade-off parameter that controls the balance between effectiveness and robustness during training.

\subsection{Natural ranking loss}
The standard training of NRMs primarily emphasizes the model's effectiveness on the labeled dataset \cite{dai2018convolutional, dai2019deeper}.
In line with existing research, we adopt a pairwise loss as the natural ranking loss, i.e., 
\begin{equation}
\label{natural loss}
    \begin{aligned}
        \mathcal{L}_{\mathrm{nat}}=-\frac{1}{|N_{q}|} \sum_{i=1}^{|N_{q}|} \log \frac{e^{f\left(\boldsymbol{q}_{i}, \boldsymbol{d}^{+}\right)}}{e^{f\left(\boldsymbol{q}_{i}, \boldsymbol{d}^{+}\right)}+\sum_{j=1}^{\bar{n}} e^{f\left(\boldsymbol{q}_{i}, \boldsymbol{d}_{j}^{-}\right)}},
    \end{aligned}
\end{equation}
where $N_{q}$ is the number of training queries, $\boldsymbol{d}^{+}$ is the relevant document and $\boldsymbol{d}^{-}$ is the irrelevant document.
We use the negative examples returned from the retrieval stage as hard negative examples and also incorporate random negative examples for the same purpose \cite{nogueira2019passage, ma2021prop}.

\subsection{Adversarial ranking loss}

To enhance adversarial robustness, we first use the WSRA attack to generate adversarial examples. 
Subsequently, we utilize augmented adversarial examples to optimize the proposed perturbation-invariant ranking loss. 

\heading{Adversarial examples} To execute a WSRA attack in a decision-based black-box  setting, \citet{wu2022prada} introduce a pseudo-relevance based adversarial ranking attack method to generate adversarial examples. 
Following this work, for each query $\boldsymbol{q}$, given a candidate document set $\mathcal{D}$, we conduct the attack against a portion of the documents evenly to derive the adversarial examples $\mathcal{D}_{adv}$.  
Each adversarial example $\boldsymbol{d}_{adv}$ in $\mathcal{D}_{adv}$ is selected from the neighborhood of the original document $\boldsymbol{d}$, based on the most threatening attack effect, i.e.,
\begin{equation}
    \begin{aligned}
        \boldsymbol{d}_{adv}= \arg \max_{\boldsymbol{d}^{\prime} \in \mathbb{B}(\boldsymbol{d}, \epsilon) } \left(f(\boldsymbol{q},\boldsymbol{d}^{\prime}) - f(\boldsymbol{q},\boldsymbol{d})\right).
    \end{aligned}
\end{equation}
Thus, we obtain adversarial examples $\mathcal{D}_{adv}$ for each query $\boldsymbol{q}$, which will be used in the following loss.

\heading{Perturbation-invariant ranking loss} As the regularization term in Eq. \ref{overall-loss}, the adversarial ranking loss encourages the model's output to be smooth, effectively constraining the sample instances within adjacent ranking decision boundaries of the model. This is achieved by minimizing the ranking order variance between the prediction of natural documents $\mathcal{D}$ and that of adversarial examples $\mathcal{D}_{adv}$. 
We design the perturbation-invariant ranking loss between $\mathcal{D}$ and $\mathcal{D}_{adv}$ as the adversarial ranking loss, i.e., 
\begin{equation}
    \label{L_adv}
    \begin{aligned}
        \mathcal{L}_{\mathrm{adv}}=-\frac{1}{|N_{q}|} \sum_{i=1}^{|N_{q}|} \psi\left(f\left(\boldsymbol{q}_{i},\mathcal{D}\right) , f\left(\boldsymbol{q}_{i},\mathcal{D}_{adv}\right)\right),
    \end{aligned}
\end{equation}
where $f(\boldsymbol{q}, \mathcal{D})$ is the predicted ranked list by a ranking model $f$ over  $\mathcal{D}$; $\psi(\cdot)$ is a differential metric to evaluate the difference in the resulting document rankings between $\mathcal{D}$ and $\mathcal{D}_{adv}$.
Here, $\mathcal{D}_{adv}$ comprises $N_{adv}$ perturbed documents and $N_{d} - N_{adv}$ benign documents.

We consider three ways to compute the difference $\psi(\cdot)$ in the ranked results obtained using $\mathcal{D}$ and $\mathcal{D}_{adv}$.

\begin{enumerate}[label=(\arabic*),wide,nosep]
\item \textbf{KL divergence.} To promote smoothness between $\mathcal{D}$ and $\mathcal{D}_{adv}$ during optimization, our objective is to minimize the KL divergence between the similarity distributions of the ranking model $f$. 
As a result, the computation of $\mathcal{L}_{\mathrm{adv}}$ in Eq. \ref{L_adv} using the $\operatorname{KL}$ divergence, is as follows:
\begin{equation}
    \begin{split}
        \mathcal{L}_{\mathrm{adv}}^{\operatorname{KL}} = & \operatorname{KL}(P(\mathcal{S} 
        \mid \mathcal{Q}, \mathcal{D}; f) \ \| \ P(\mathcal{S} \mid \mathcal{Q}, \mathcal{D}_{adv}; f))\\
        = & \frac{1}{N_{q}} \sum_{i=1}^{N_{q}} \sum_{j=1}^{N_{d}} P(s_{i} \mid \boldsymbol{q}_{i}, \boldsymbol{d}_{j} \sim \mathcal{D}; f) \cdot{} \\
        & \quad \log \frac{P(s_{i} \mid \boldsymbol{q}_{i}, \boldsymbol{d}_{j} \sim \mathcal{D}; f) }{P(s_{i} \mid \boldsymbol{q}_{i}, \boldsymbol{d}^{\prime}_{j} \sim \mathcal{D}_{adv}; f)}, \\
    \end{split}
\end{equation}
where 
\begin{equation*}
    \begin{aligned}
        P(s_{i} \mid \boldsymbol{q}_{i}, \boldsymbol{d}_{j} \sim \mathcal{D}; f) 
        &= \frac{\exp \! \left(f\left(\boldsymbol{q}_{i}, \boldsymbol{d}_{j}\right)\right)}{\sum_{\boldsymbol{d}_{k} \sim \mathcal{D}} \exp \! \left(f\left(\boldsymbol{q}_{i}, \boldsymbol{d}_{k}\right)\right)},
        \\
%
        P(s_{i} \mid \boldsymbol{q}_{i}, \boldsymbol{d}^{\prime}_{j} \sim \mathcal{D}_{adv}; f) 
        &= \frac{\exp \! \left(f\left(\boldsymbol{q}_{i}, \boldsymbol{d}^{\prime}_{j}\right)\right)}{\sum_{\boldsymbol{d}^{\prime}_{k} \sim \mathcal{D}_{adv}} \exp \! \left(f\left(\boldsymbol{q}_{i}, \boldsymbol{d}^{\prime}_{k}\right)\right)}.
    \end{aligned}
\end{equation*}
Let us consider a scenario where only one document ranked at the bottom within $\mathcal{D}$ is perturbed and moves to the top-1 position, while the other documents are shifted down one position each.
In this case, the distribution of the entire permutation would not undergo significant disordering.
However, even though the overall re-ordering might be limited, the situation could have implications for practical search engines.
Therefore, using KL divergence as a metric may not impose a sufficiently severe penalty for this attack result.

Next, we present alternatives to tackle this issue. 
We introduce a listwise loss 
to model the output ranking both before and after perturbation. 
By concentrating on the ranked list, our approach strives to prevent the perturbed document from excessively rising to the top position, thereby preserving a natural and gradual change in rankings.
\item \textbf{Listwise function -- ListNet.} 
ListNet \cite{cao2007learning} devises a listwise loss to assess the dissimilarity between the predicted ranked list and the ground-truth permutation, given by the following expression: 
\begin{equation}
    \mbox{}\hspace*{-2mm}
    \begin{aligned}
        \mathcal{L}_{\operatorname{ListNet}}&\left(f ; \boldsymbol{q}, \mathcal{D}, \mathcal{Y} \right)= {}
        \\
        &\operatorname{KL} \left( P ( \pi_{f} \mid \varphi (f(\boldsymbol{q},\mathcal{D}))) \ \| \ P(\pi_{\mathcal{Y}}) \right),
    \end{aligned}
\end{equation}
where $\pi_{f}$ is the permutation predicted by $f$, $\pi_{\mathcal{Y}}$ is the ground-truth permutation, and $\varphi$ is a transformation function (an increasing and strictly positive function, e.g., linear, exponential or sigmoid).  
The probability of a permutation given the score list \cite{cao2007learning}, is computed as follows,  
\begin{equation}
    \begin{aligned}
        P(\pi_{f} \mid  \varphi (f(\boldsymbol{q},\mathcal{D}))) =\prod_{j=1}^{N_{d}} \frac{\varphi\left(f_{\pi(j)} (\boldsymbol{q}, \mathcal{D}) \right)}{\sum_{k=j}^{N_{d}} \varphi\left(f_{\pi(k)}(\boldsymbol{q}, \mathcal{D})\right)},
    \end{aligned}
\end{equation}
where $f_{\pi(i)}(\boldsymbol{q}, \mathcal{D}) $ denotes the similarity score predicted by $f$ of the document, which is ranked at the $i$-th position with respect to the query $\boldsymbol{q}$. 
We define $\mathcal{L}_{adv}$ based on ListNet as, 
\begin{equation}
    \mbox{}
    \hspace*{-1.75mm}
    \begin{aligned}
        & \mathcal{L}_{\mathrm{adv}}^{\operatorname{ListNet}}  =  \\
        & \operatorname{KL} \left(P(\pi_{f 
        (\boldsymbol{q}, \mathcal{D})} \mid f(\boldsymbol{q},\mathcal{D}_{adv})) \| P(\pi_{f (\boldsymbol{q}, \mathcal{D})} \mid f(\boldsymbol{q},\mathcal{D}) )\right),
        \hspace*{-2.5mm}\mbox{}
    \end{aligned}
\end{equation}
where $ \pi_{f (\boldsymbol{q}, \mathcal{D})} $ is the permutation computed by the ranking model $f$ on data pair $(\boldsymbol{q}, \mathcal{D})$.

\item \textbf{Listwise function -- ListMLE.}
ListMLE \cite{xia2008listwise} addresses the computational complexity of ListNet  by optimizing the negative log-likelihood of the ground-truth permutation $\pi_{\mathcal{Y}}$, i.e.,  
\begin{equation}
    \begin{aligned}
        \mathcal{L}_{\operatorname{ListMLE}}\left(f ; \boldsymbol{q}, \mathcal{D}, \mathcal{Y} \right)= -\log P(\pi_{\mathcal{Y}} \mid f(\boldsymbol{q},\mathcal{D})).
    \end{aligned}
\end{equation}
Inspired by ListMLE, we design our adversarial loss $\mathcal{L}_{adv}$ to compute the negative log-likelihood of benign document permutation, i.e.,
\begin{equation}
    \begin{aligned}
        \mathcal{L}_{\mathrm{adv}}^{\operatorname{ListMLE}}= -\log P(\pi_{f (\boldsymbol{q}, \mathcal{D})} \mid f(\boldsymbol{q},\mathcal{D}_{adv})),
    \end{aligned}
\end{equation}
where $\pi_{f (\boldsymbol{q}, \mathcal{D})}$ represents the document permutation generated by the ranking model $f$ for the list of documents that have not been attacked. 
This enables us to effectively align the ranked list after perturbations with the benign ranked list, thereby achieving adversarial robustness. 
\end{enumerate}

\vspace{-1mm}
\section{Experiments}
We present our experimental setup and results in this section.

\begin{table*}[t]
\small
\centering
   \vspace{-3mm}
  	\begin{tabular}{l l  c c c  c c c}
  \toprule
   Model & Method & CleanMRR@10 & CleanMRR@100 & RobustMRR@10 & RobustMRR@100 & ASR$\downarrow$ & LSD$\downarrow$ \\
       \midrule
BM25 & -  & 0.1874 & 0.1985 & 0.1624 & 0.1736 & 56.4 & 15.3 \\
\midrule
    \multirow{7}{*}{ConvKNRM} 
& ST  & 0.2461 & 0.2592 & 0.1692 & 0.1741 & 95.1 & 33.2\\
& DA  & 0.2298 & 0.2378 & 0.1786 & 0.1829 & 71.2 & 23.6\\
& CertDR  & 0.1816 & 0.1935 & 0.1592 & 0.1632 & 65.3 & 19.3\\
& AT  & 0.2316 & 0.2410 & 0.1896 & 0.1953 & 61.3 & 17.6\\
 \cmidrule{2-8}
& PIAT$_{\operatorname{KL}}$  & 0.2498 & 0.2603 & 0.2008\rlap{$^{\ast}$} & 0.2073\rlap{$^{\ast}$} & 51.1\rlap{$^{\ast}$} & 12.6 \\
& PIAT$_{\operatorname{ListNet}}$  & 0.2513\rlap{$^{\ast}$} & 0.2621\rlap{$^{\ast}$} & \textbf{0.2035}\rlap{$^{\ast}$} & \textbf{0.2009}\rlap{$^{\ast}$} & \textbf{48.3}\rlap{$^{\ast}$} & \textbf{11.5}\rlap{$^{\ast}$} \\
& PIAT$_{\operatorname{ListMLE}}$  & \textbf{0.2534}\rlap{$^{\ast}$} & \textbf{0.2645}\rlap{$^{\ast}$} & 0.2018\rlap{$^{\ast}$} & 0.2091\rlap{$^{\ast}$} & 49.2\rlap{$^{\ast}$} & 11.9 \\
 \midrule
    \multirow{7}{*}{BERT} 
& ST  & 0.3831 & 0.3923 & 0.3225 & 0.3286 & 92.1 & 32.3\\
& DA  & 0.3705 & 0.3810 & 0.3315 & 0.3423 & 63.2 & 18.9 \\
& CertDR  & 0.3202 & 0.3311 & 0.3026 & 0.3140 & 56.9 & 16.3\\
& AT  & 0.3743 & 0.3865 & 0.3451 & 0.3508 & 55.1 & 15.6\\
 \cmidrule{2-8}
& PIAT$_{\operatorname{KL}}$  & 0.3860 & 0.3948 & 0.3686\rlap{$^{\ast}$} & 0.3761\rlap{$^{\ast}$} & 41.2\rlap{$^{\ast}$} & 9.4 \\
& PIAT$_{\operatorname{ListNet}}$  & 0.3892 & 0.3981 & \textbf{0.3728}\rlap{$^{\ast}$} & \textbf{0.3802}\rlap{$^{\ast}$} & \textbf{36.1}\rlap{$^{\ast}$} & \textbf{7.2}\rlap{$^{\ast}$} \\
& PIAT$_{\operatorname{ListMLE}}$  & \textbf{0.3910}\rlap{$^{\ast}$} & \textbf{0.4002}\rlap{$^{\ast}$} & 0.3705\rlap{$^{\ast}$} & 0.3785\rlap{$^{\ast}$} & 38.3\rlap{$^{\ast}$} & 7.9 \\
 \midrule
    \multirow{7}{*}{PROP} 
& ST  & 0.3902 & 0.4061 & 0.3352 & 0.3478 & 90.3 & 30.9 \\
& DA  & 0.3783 & 0.3930 & 0.3418 & 0.3538 & 60.4 & 16.8 \\
& CertDR  & 0.3351 & 0.3489 & 0.3199 & 0.3220 & 52.8 & 13.4 \\
& AT  & 0.3819 & 0.4002 & 0.3532 & 0.3611 & 51.2 & 12.8 \\
 \cmidrule{2-8}
& PIAT$_{\operatorname{KL}}$  & 0.3943 & 0.4063 & 0.3749\rlap{$^{\ast}$} & 0.3853\rlap{$^{\ast}$} & 39.4\rlap{$^{\ast}$} & 8.2 \\
& PIAT$_{\operatorname{ListNet}}$  & 0.3971 & 0.4121 & \textbf{0.3794}\rlap{$^{\ast}$} & \textbf{0.3890}\rlap{$^{\ast}$} & \textbf{35.0}\rlap{$^{\ast}$} & \textbf{6.2}\rlap{$^{\ast}$} \\
& PIAT$_{\operatorname{ListMLE}}$  & \textbf{0.3992}\rlap{$^{\ast}$} & \textbf{0.4148}\rlap{$^{\ast}$} & 0.3767\rlap{$^{\ast}$} & 0.3864\rlap{$^{\ast}$} & 37.8\rlap{$^{\ast}$} & 7.8 \\
\bottomrule
    \end{tabular}
       \caption{Trade-off performance of different ranking models under PIAT and defense baselines; For CertDR, the ASR is evaluated under conditional success rate \cite{wu2022certified}; $\ast$ indicates significant improvements over the best baseline ($p \le 0.05$).}
   \label{table:Baseline}
   \vspace{-6mm}
\end{table*}

\vspace{-1mm}
\subsection{Experimental setup}
\textbf{Dataset and target ranking models.}
We conduct experiments on the \emph{MS MARCO Passage Ranking dataset}, which is a large-scale benchmark dataset for Web passage retrieval, with about 8.84 million passages  \cite{nguyen2016ms}. 
The relevant documents to user queries are obtained using Bing, thereby simulating real-world web search scenarios. 

We choose several typical ranking models that achieve promising effectiveness, including traditional probabilistic models, e.g., \emph{BM25} \cite{robertson1994some},  interaction-focused NRMs, e.g., \emph{ConvKNRM} \cite{Dai_Xiong_Callan_Liu_2018}, and pre-trained models, e.g., \emph{BERT} \cite{Devlin_Chang_Lee_Toutanova_2019} and \emph{PROP} \cite{ma2021prop}, for adversarial attack. 

\heading{Evaluation metrics} 
\begin{enumerate*}[label=(\roman*)]
    \item \emph{CleanMRR@$k$} evaluates Mean Reciprocal Rank (MRR) performance on the clean dataset \cite{ma2021b, yan2021unified}.  
    \item \emph{RobustMRR@$k$} evaluates the MRR performance on the attacked dataset by WSRA. 
    \item \emph{Attack success rate (ASR)} (\%) evaluates the percentage of the after-attack documents that are ranked higher than original documents \cite{wu2022prada}.  
    \item \emph{Location square deviation (LSD)} (\%) evaluates the consistency between the original  and perturbed ranked list for a query, by calculating the average deviation between the document positions in the two lists \cite{sun2022reorder}. 
\end{enumerate*}

The effectiveness of a ranking model is better with a higher CleanMRR. 
The robustness of a ranking model is better with a higher RobustMRR and a lower ASR and LSD. 

\heading{Baselines} 
\begin{enumerate*}[label=(\roman*)]
    \item \emph{Standard training (ST)}: We directly optimize the ranking model via the natural ranking loss (Eq. \ref{natural loss}) without defense mechanisms. 
    \item \emph{Data augmentation (DA)}: We augment each document in the collection with $2$ new documents by uniformly replacing synonyms, and then use the normal hinge loss for training following \cite{wu2022certified}. The number of replacement words equals the number of words perturbed by the WSRA attack. 
    \item \emph{Adversarial training (AT)}: We follow the vanilla AT method  \cite{IanGoodfellow2014ExplainingAH} to directly include  the adversarial examples during training. 
    \item \emph{CertDR} is a certified defense method for NRMs \cite{wu2022certified}, which achieves certified top-$K$ robustness against WSRA attacks.  
\end{enumerate*}

\heading{Implementation details}
We implement target ranking models following previous work~\cite{Dai_Xiong_Callan_Liu_2018, Devlin_Chang_Lee_Toutanova_2019, ma2021prop, liu2023black}.
First-stage retrieval is performed using the Anserini toolkit \cite{yang2018anserini} with BM25, to obtain top 100 candidate passages. 
The ranked list is obtained by using the well-trained ranking model to re-rank the above initial candidate set.  

We randomly sample 1000 Dev queries as target queries to attack their ranked lists for evaluation.  
For each sampled query, we randomly sample 1 document from 9 ranges in the ranked list following \cite{wu2022prada}, i.e., $[11,20], ..., [91,100]$, respectively. 
We attack these 9 target documents to achieve their corresponding adversarial examples using WSRA. 
Finally, we evaluate the defense performance of ranking models using the attacked list with 9 adversarial examples and its query as an input. 
For BM25, we attack it using adversarial examples generated by the attack method in \cite{wu2022prada} designed for attacking BERT. 

For adversarial training, considering the time overhead, we sample 0.1 million (1/10 of the total) training queries to generate adversarial examples.
For each training query, we randomly sample 10 documents from its initial candidate set to construct adversarial examples using WSRA. 
Note the sampled documents are not ground-truth ones. 
We set the maximum number of word substitutions to 20, and other hyperparameters are consistent with \citet{wu2022prada}.
The regularization hyperparameter $\lambda$ is set to 0.5.
We train the NRMs with a batch size of 100, maximum sequence length of 256, and learning rate of 1e-5.

By training the ranking model with different adversarial ranking losses, i.e., $\mathcal{L}_{\mathrm{adv}}^{\operatorname{KL}}$, $\mathcal{L}_{\mathrm{adv}}^{\operatorname{ListNet}}$, and $\mathcal{L}_{\mathrm{adv}}^{\operatorname{ListMLE}}$, 
we obtain three  types of PIAT as \emph{PIAT$_{\operatorname{KL}}$}, \emph{PIAT$_{\operatorname{ListNet}}$}, and \emph{PIAT$_{\operatorname{ListMLE}}$}, respectively. 

\vspace{-1mm}
\subsection{Experimental results}
\heading{Defense comparison}  Table~\ref{table:Baseline} presents a comparison of the trade-off performance among four ranking models with different defenses.  
Observations on the defense baselines are: 
\begin{enumerate*}[label=(\roman*)]
    \item Effectiveness and adversarial robustness of PROP is generally better than BERT, which in turn is stronger than ConvKNRM. 
    This indicates that well-designed model architectures and pre-training objectives encourage a ranking model to achieve better trade-off performance.
    \item After being attacked, the ranking performance of the ST method without defense mechanisms, decreases significantly with a high ASR and LSD. 
    Hence, it is imperative not only to focus on the effectiveness of existing NRMs when deploying them in real-world scenarios.
    \item CertDR ensures consistent ranking performance between clean and adversarial data. 
    This could be attributed to CertDR's ability to guarantee the stability of the Top-$K$ of the ranked list by certifying the Top-$K$ robustness. 
    \item DA and AT enhance the model's ranking performance on adversarial data, but this improvement comes at the cost of reduced performance on clean data. 
    The finding is consistent with prior research in natural language processing and machine learning \cite{zhang2019theoretically,rade2021helper, bao2021defending}. 
\end{enumerate*}

When we look at PIAT, we find that:
\begin{enumerate*}[label=(\roman*)]
    \item In general, three types of PIAT exhibit superior effectiveness and adversarial robustness than baselines. 
    This suggests that a combination of proposed supervised and semi-supervised training enables the effective utilization of information from extensive unlabeled documents to enhance trade-off performance.
    \item PIAT outperforms the baselines in terms of LSD, indicating increased resistance to perturbations across the entire ranked list. This highlights the efficacy of the perturbation-invariant ranking loss in facilitating NRMs to learn more robust ranking decision boundaries. 
    \item PIAT$_{\operatorname{KL}}$ demonstrates comparatively lower effectiveness in comparison to the other two PIAT types, likely due to the fact that the KL divergence of the relevant scores serves as a soft constraint, rendering a relatively mild supervisory signal. 
    \item PIAT$_{\operatorname{ListMLE}}$ achieves a slightly inferior performance compared to PIAT$_{\operatorname{ListNet}}$. 
    The reason might be that ListMLE compromises precision by converting list-wise differences into an estimate of the probability distribution. Nevertheless, ListMLE exhibits higher training efficiency.
\end{enumerate*}

\begin{figure}[t]
    \centering
    \includegraphics[width=\linewidth]{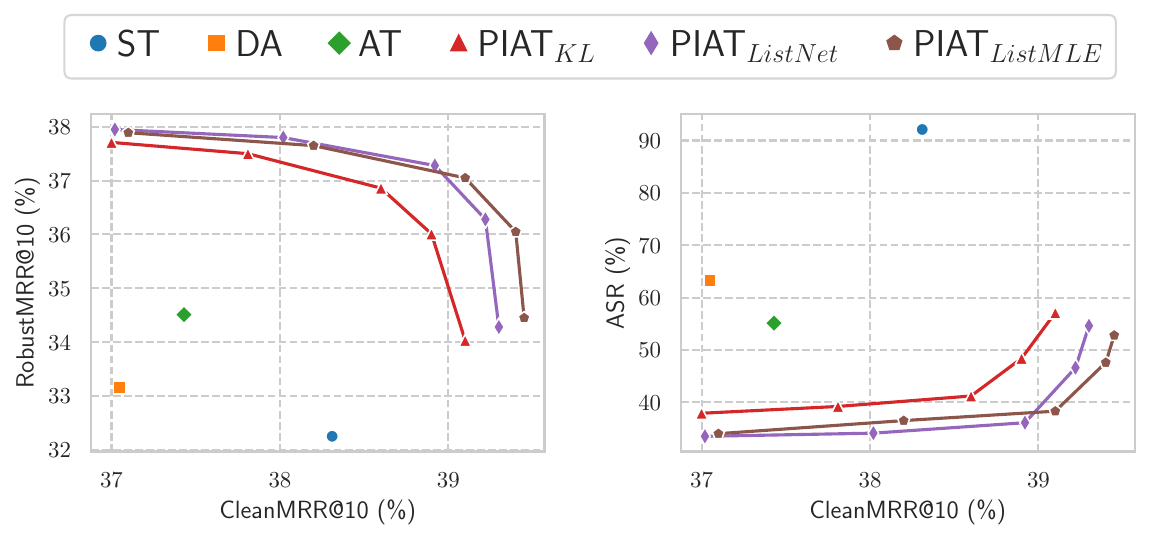}
    \vspace*{-6mm}
    \caption{The sensitivity of trade-off parameter $\lambda$ exhibited by different types of PIAT, compared with empirical defense methods. From left to right, we increase the trade-off parameter $\lambda$ of PIAT from 0.2 to 0.8 with the step of 0.15.}
    \label{fig:tradeoff}
    \vspace{-6mm}
\end{figure}

\heading{Effectiveness vs. robustness trade-off} $\lambda$ is an important hyperparameter in our proposed method, since it plays a crucial role in determining the balance between effectiveness and robustness.
Figure \ref{fig:tradeoff} shows a comparison of the effectiveness vs.\ robustness trade-off between PIAT and empirical defense baselines. 
CertDR is excluded from the comparison due to its inferior performance compared to empirical defenses in terms of both effectiveness and robustness, as indicated in Table~\ref{table:Baseline}. 
We conduct comparisons by examining CleanMRR@10 (for effectiveness) against RobustMRR@10 and ASR (for robustness) respectively, thereby visualizing the effectiveness-robustness trade-off.

We show the results of the BERT model; similar findings were obtained for other ranking models.
Both DA and AT enhance robustness, but at the expense of effectiveness.
This suggests they may not adequately consider the balanced relationship between effectiveness and robustness.
When we look at the different types of PIAT we find that they achieve a heightened trade-off between effectiveness and robustness.
This indicates that proper modeling and optimization of the boundary ranking error can guide NRMs to bolster robustness, while maintaining or even improving effectiveness.
Furthermore, we note that with an excessively large $\lambda$, the model's robustness considerably decreases, while effectiveness exhibits marginal growth. 
Conversely, an excessively small $\lambda$ shows a notable decline in effectiveness, while robustness experiences only a minor improvement. 
These findings emphasize the necessity of prioritizing the balance of effectiveness and robustness when training NRMs.

\begin{figure}[t]
    \centering
    \includegraphics[width=\linewidth]{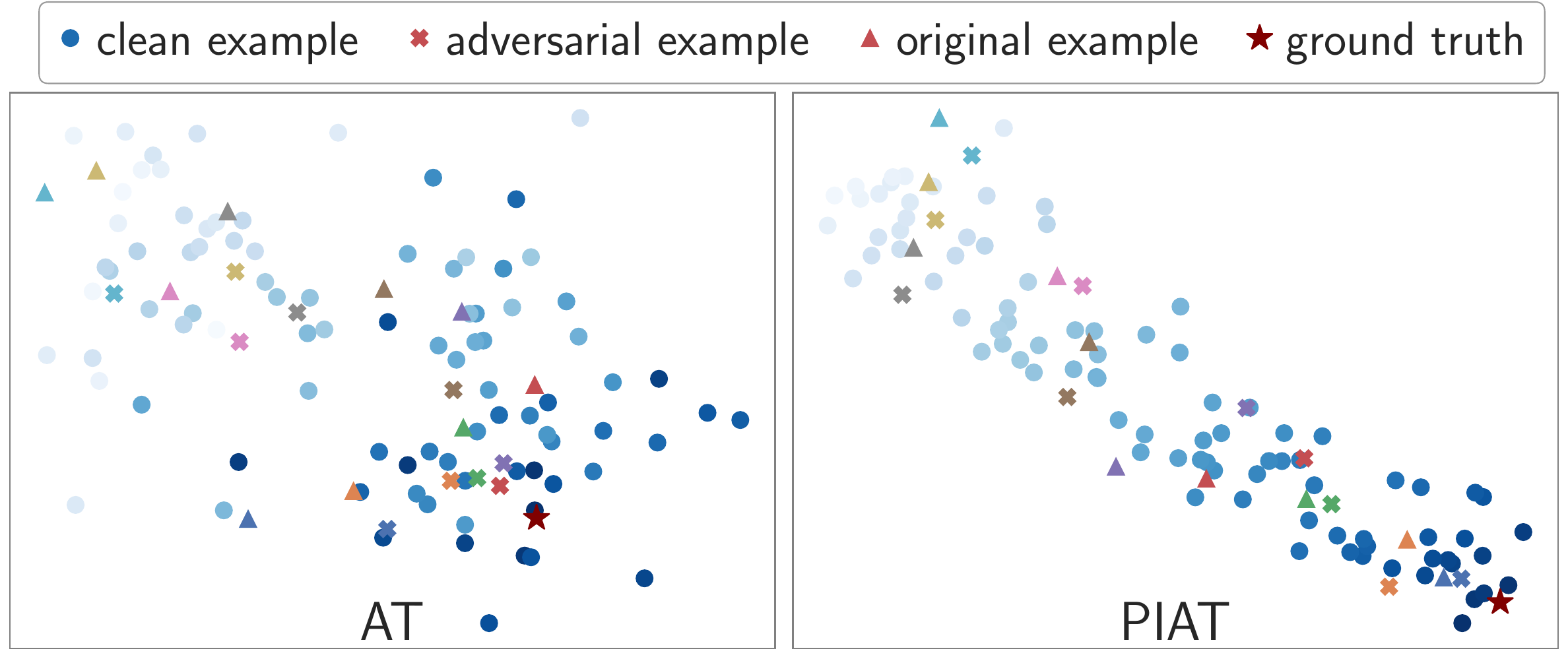}
    \vspace*{-6mm}
    \caption{t-SNE plot of query-document representations for AT and PIAT. For clean examples, a darker blue circle represents a higher relevance score. Triangle and cross denote the original document and its corresponding adversarial example. Star denotes the ground-truth query-document pair.}
    \label{fig:tsne}
    \vspace{-7mm}
\end{figure}

\heading{Visual analysis} We train the BERT model using AT and PIAT$_\mathit{ListMLE}$, respectively, on the MS MARCO passage ranking dataset, with inputs being query-document concatenations.
The hidden states of [CLS] in BERT's final layer are utilized as query-document pair representations and visualized using t-SNE \cite{van2008visualizing} to observe semantic space distributions. 
As a t-SNE example, we generate plots by sampling a query (QID=262232) and selecting its top 100 candidate documents, including 9 adversarial examples. 
Results in Figure \ref{fig:tsne} show that: 
\begin{enumerate*}[label=(\roman*)]
\item For AT, the distribution of adversarial examples in the latent space is relatively disordered. 
By examining data at the same relevance level (color depth), we find that the decision boundaries exhibit a certain degree of chaos. 
Some adversarial examples have managed to move away from their original positions and closer to the ground-truth. 
AT lacks clear distinctions in terms of modeling effectiveness and robustness, relying solely on pair-wise loss for simultaneous optimization.
\item For PIAT, the ranking decision boundary not only distinguishes between data points of varying relevance levels, but also effectively constrains the adversarial examples to stay close to their original examples. 
This result emphasizes the fact that by using perturbation-invariant loss, tailored through analysis of boundary ranking errors, PIAT achieves remarkable adversarial robustness while maintaining effectiveness compared to the traditional AT. 
\end{enumerate*}
Similar observations were obtained with other ranking models.

\vspace{-2mm}
\section{Related Work}

\textbf{Neural ranking models.}
The emergence of deep learning has led to the popularity of NRMs \cite{KezbanDilekOnal2018NeuralIR,guo2020deep}, showcasing their superiority over traditional ranking models. 
There have been efforts to leverage pre-trained models for ranking tasks \cite{fan2022pre}, further enhancing the effectiveness of NRMs. 
Additionally, studies have explored training NRMs using data augmentation techniques, such as hard negative mining \cite{Xiong2021ApproximateNN,Zhan2021OptimizingDR}, achieving new state-of-the-art performance. 
Despite these effectiveness improvements, these studies often overlook the adversarial robustness of NRMs.

\heading{Defense methods}
In recent years, model robustness has attracted attention in various fields \cite{wang2021discover,Liangke_SymCLKG_TKDE,wang2022micro,liang2023learn}, including IR \cite{liu2023black,liu2023robustness}.
In the context of robustness, adversarial attacks aim to discover human-imperceptible perturbations that can deceive neural networks \cite{szegedy2014intriguing}. 
\citet{wu2022prada} introduced the WSRA method of attacking black-box NRMs using word substitution. 
This study revealed the serious vulnerability of NRMs to synonym substitution perturbations. 
As a result, subsequent explorations of attack against NRMs have emerged \cite{liu2023topicoriented,liu2022order,chen2023towards}, inspired by this pioneering work.

In response to adversarial attacks, research has proposed various defense strategies to enhance adversarial robustness. 
These can be generally classified into \emph{certified defenses} and \emph{empirical defenses}. 
Certified defenses aim for theoretical robustness against specific adversarial perturbations \cite{raghunathan2018certified}. 
For instance, \citet{wu2022certified} introduced a certified defense method that ensures the top-K robustness of NRMs via randomized smoothing. 
However, due to their theoretical nature, these methods often face limitations in practical applications and may not fully meet the desired performance requirements. 

Empirical defenses aim to enhance the empirical robustness of models against known adversarial attacks, and this approach has been extensively explored in  image classification \cite{madry2018towards,wang2019convergence} and text classification~\cite{ye2020safer, jia2019certified}. 
Among these methods, adversarial training emerges as one of the most effective defenses. 
Adversarial training on adversarial examples remains empirically robust \cite{cui2021learnable}. 
However, the use of adversarial training as a defensive mechanism is often limited to simple  classification scenarios, and its application in NRMs remains largely unexplored. 
Therefore, we propose an adversarial training method tailored for NRMs to improve the trade-off between effectiveness and robustness.

\vspace{-5mm}
\section{Conclusion}

To the best of our knowledge, our study is the first study on the trade-off between effectiveness and adversarial robustness for neural retrieval models. 
Our theoretical analysis motivated the development of perturbation-invariant adversarial training, incorporating a new regularized surrogate loss. 
Experimental results have showcased the superior performance of our method in terms of effectiveness and robustness.

\heading{Broader impact and limitations} 
We aim to make an initial exploration for adversarial robustness and to inspire the IR community to further enhance the effectiveness-robustness trade-off. 
As to the limitations of our work, we currently only consider the popular attack of WSRA, and constructing adversarial training examples could be time-consuming.
In future work, we will investigate the design of adversarial training methods to defend against other or unseen attacks, and create training examples with reduced time overhead.

\bibliography{references}

\clearpage
\appendix
\section{Appendix}
\subsection{Proof of Theorem 1} \label{proof:1}

According to Definition 3.3, the boundary ranking error is characterized as the proportion of examples that were initially ranked accurately but became misranked because of their proximity to the ranking model's boundary. 
In order to mitigate this boundary ranking error, we propose the incorporation of a trainable upper bound as an alternative. This measure guarantees an improved robustness of the ranking model, effectively tackling vulnerabilities associated with the boundary ranking error.

To derive a trainable upper bound, we employ the neighborhood of $(\boldsymbol{q}, \boldsymbol{d}_{i})$, denoted as $\pi_{n}(\boldsymbol{q}, \boldsymbol{d}_{i}) = \pi_{y}(\boldsymbol{q}, \boldsymbol{d}_{i}) - 1$. 
This serves as a bridge to determine the objective for unsupervised documents in relation to the perturbed document $\boldsymbol{d}_{i}$. Specifically, the boundary ranking error is calculated by, 
\setcounter{equation}{17} 
\begin{equation}
\mbox{}\hspace*{-1mm}
\begin{split}
    \mathcal{R}&_{\mathrm{bdy}}(f) \\
    & = \mathbb{P}[ \boldsymbol{d}_{i} \in \mathbb{B}(\text{DB}(f), \epsilon) , \pi_{f}(\boldsymbol{q}, \boldsymbol{d}_{i}) = \pi_{y}(\boldsymbol{q}, \boldsymbol{d}_{i})  ] ,
    \hspace*{-4mm}\mbox{}
    \\ 
    & \leq  \mathbb{P}[ \boldsymbol{d}_{i} \in \mathbb{B}(\text{DB}(f), \epsilon)  ], \\
    & = \mathbb{E}_{\boldsymbol{d}_{i} \sim \mathcal{D}} \max _{\boldsymbol{d_{i}}^{\prime} \in \mathbb{B}(\boldsymbol{d_{i}}, \epsilon)} \mathbb{I} \{ [ \pi_{f}(\boldsymbol{q}, \boldsymbol{d}_{i}) \!-\! \pi_{n}(\boldsymbol{q}, \boldsymbol{d}_{i})   ] \\ 
    & \qquad \qquad \qquad  \cdot [ \pi_{f}(\boldsymbol{q}, \boldsymbol{d}^{\prime}_{i}) \!-\! \pi_{n}(\boldsymbol{q}, \boldsymbol{d}_{i})] \!\leq\! 0 \} , \\
    &  =\mathbb{E}_{\boldsymbol{d}_{i} \sim \mathcal{D}} \max _{\boldsymbol{d_{i}}^{\prime} \in \mathbb{B}(\boldsymbol{d_{i}}, \epsilon)} \mathbb{I} \{ [ \pi_{f}(\boldsymbol{q}, \boldsymbol{d}_{i}) - \pi_{n}(\boldsymbol{q}, \boldsymbol{d}_{i})   ] \\
    & \qquad \qquad \qquad  \neq [ \pi_{f}(\boldsymbol{q}, \boldsymbol{d}^{\prime}_{i}) - \pi_{n}(\boldsymbol{q}, \boldsymbol{d}_{i})] \}, \\
    &  = \mathbb{E}_{\boldsymbol{d}_{i} \sim \mathcal{D}} \max _{\boldsymbol{d}_{i}^{\prime} \in \mathbb{B}(\boldsymbol{d}_{i}, \epsilon)} \mathbb{I}\left\{ \pi_{f}(\boldsymbol{q}, \boldsymbol{d}_{i})  \neq \pi_{f}(\boldsymbol{q}, \boldsymbol{d}^{\prime}_{i}) \right\}.
    \hspace*{-4mm}\mbox{}
\end{split}
\end{equation}
This completes the proof of Theorem 1. 
This showcases the possibility of refining the upper bound for the boundary ranking error without necessitating dependence on the ground-truth ranking list $\pi_{y}$. Instead, we have devised an objective aimed at promoting the invariance of the ranked list against perturbations, thereby giving rise to the loss function presented in Eq. 11.

\subsection{Tightness of Theorem 1} 
\label{proof:2}
Whether the upper bound in Theorem 1 is sufficiently tight is of great importance. In the following, we provide the necessary proof to state its tightness. 
First, we present the gap between the upper bound of the boundary ranking error and the error itself. 
Then, we prove that this gap has an upper limit, which is closely related to the natural ranking error. 
Finally, by demonstrating that the upper limit of the gap can be effectively optimized, we prove that the upper bound of the boundary ranking error in Theorem 1 is sufficiently tight.

\heading{The gap between the boundary ranking error and its upper bound} Since the upper bound of the boundary ranking error is always larger than itself, the gap $\mathcal{G}$ between the boundary ranking error $\mathcal{R}_{\mathrm{bdy}}(f)$ and its upper bound $\mathcal{R}^*_{\mathrm{bdy}}(f)$ is denoted as, 
\begin{equation}
\label{tigeness gap}
\mbox{}\hspace*{-1mm}
\begin{split}
\mathcal{G} & \left(\mathcal{R}^*_{\mathrm{bdy}}(f), \mathcal{R}_{\mathrm{bdy}}(f) \right) \\
& = \left|\mathcal{R}^*_{\mathrm{bdy}}(f) - \mathcal{R}_{\mathrm{bdy}}(f) \right|  \\
& = \mathcal{R}^*_{\mathrm{bdy}}(f) - \mathcal{R}_{\mathrm{bdy}}(f) \\
& = \mathcal{R}^*_{\mathrm{bdy}}(f) \\
& - \mathbb{P}[ \boldsymbol{d}_{i} \in \mathbb{B}(\text{DB}(f), \epsilon) , \pi_{f}(\boldsymbol{q}, \boldsymbol{d}_{i}) = \pi_{y}(\boldsymbol{q}, \boldsymbol{d}_{i}) ],\\
\end{split}
\end{equation}
where $\pi_{f}(\boldsymbol{q}, \boldsymbol{d}_{i})$ and $\pi_{y}(\boldsymbol{q}, \boldsymbol{d}_{i})$ are the predicted ranking position and the ground truth position, respectively.

Based on Eq. 18 in the above proof of Theorem 1,  $\mathcal{R}^*_{\mathrm{bdy}}(f)$ is denoted as, 
\begin{equation}
\mbox{}\hspace*{-1mm}
\begin{split}
\mathcal{R}^*_{\mathrm{bdy}}(f) & = \mathbb{E}_{\boldsymbol{d}_{i} \sim \mathcal{D}} \max _{\boldsymbol{d}_{i}^{\prime} \in \mathbb{B}(\boldsymbol{d}_{i}, \epsilon)} \mathbb{I}\left\{ \pi_{f}(\boldsymbol{q}, \boldsymbol{d}_{i})  \neq \pi_{f}(\boldsymbol{q}, \boldsymbol{d}^{\prime}_{i}) \right\} \\ 
& = \mathbb{P}[ \boldsymbol{d}_{i} \in \mathbb{B}(\text{DB}(f), \epsilon) ].
\label{upper bound}
\end{split}
\end{equation}
To simplify Eq.~\ref{tigeness gap}, we introduce the model predicted rankings for document $\boldsymbol{d}_{i}$ to compose a joint probability.
According to the definition in Eq. 4, the neighborhood of the ranking decision boundary for a ranking model $f$ is denoted as, 
\begin{equation}
\label{decision boundary neighborhood}
    \begin{split}
        & \mathbb{B}(\text{DB}(f), \epsilon) := \{ \boldsymbol{d}_{i} \sim \mathcal{D} : \ \exists ~\boldsymbol{d}_{i}^{\prime} \in \mathbb{B}(\boldsymbol{d_{i}}, \epsilon) \text{ s.t.} \\
        & [ \pi_{f}(\boldsymbol{q}, \boldsymbol{d}_{i}) \!-\! \pi_{n}(\boldsymbol{q}, \boldsymbol{d}_{i})   ] \cdot [ \pi_{f}(\boldsymbol{q}, \boldsymbol{d}^{\prime}_{i}) \!-\! \pi_{n}(\boldsymbol{q}, \boldsymbol{d}_{i})] \!\leq\! 0  \},
    \end{split}
\end{equation}
where $ \pi_{n}(\boldsymbol{q}, \boldsymbol{d}_{i}) = \pi_{y}(\boldsymbol{q}, \boldsymbol{d}_{i}) - 1$ is the neighborhood ranking of $(\boldsymbol{q}, \boldsymbol{d}_{i})$, with the attacker's objective being to achieve a higher ranking than it. 

According to the definition of the neighborhood of the ranking model decision boundary in Eq.~\ref{decision boundary neighborhood}, we can further expand Eq.~\ref{upper bound} as follows,
\begin{equation}
\label{}
\mbox{}\hspace*{-1mm}
\begin{split}
\mathcal{R}^*_{\mathrm{bdy}}(f) 
& = \mathbb{P}[ \boldsymbol{d}_{i} \in \mathbb{B}(\text{DB}(f), \epsilon) ] \\
& = \mathbb{P}[ \boldsymbol{d}_{i} \in \mathbb{B}(\text{DB}(f), \epsilon) , \pi_{f}(\boldsymbol{q}, \boldsymbol{d}_{i}) < \pi_{n}(\boldsymbol{q}, \boldsymbol{d}_{i}) ] \\
& + \mathbb{P}[ \boldsymbol{d}_{i} \in \mathbb{B}(\text{DB}(f), \epsilon) , \pi_{f}(\boldsymbol{q}, \boldsymbol{d}_{i}) \geq \pi_{n}(\boldsymbol{q}, \boldsymbol{d}_{i}) ]. \\
\end{split}
\end{equation}
Considering $\pi_{n}(\boldsymbol{q}, \boldsymbol{d}_{i}) = \pi_{y}(\boldsymbol{q}, \boldsymbol{d}_{i}) - 1$, we have, 
\begin{equation}
\label{last bdy}
\mbox{}\hspace*{-1mm}
\begin{split}
\mathcal{R}^*_{\mathrm{bdy}}(f) 
& = \mathbb{P}[ \boldsymbol{d}_{i} \in \mathbb{B}(\text{DB}(f), \epsilon) , \pi_{f}(\boldsymbol{q}, \boldsymbol{d}_{i}) < \pi_{n}(\boldsymbol{q}, \boldsymbol{d}_{i}) ] \\
& + \mathbb{P}[ \boldsymbol{d}_{i} \in \mathbb{B}(\text{DB}(f), \epsilon) , \pi_{f}(\boldsymbol{q}, \boldsymbol{d}_{i}) = \pi_{n}(\boldsymbol{q}, \boldsymbol{d}_{i}) ] \\
& + \mathbb{P}[ \boldsymbol{d}_{i} \in \mathbb{B}(\text{DB}(f), \epsilon) , \pi_{f}(\boldsymbol{q}, \boldsymbol{d}_{i}) = \pi_{y}(\boldsymbol{q}, \boldsymbol{d}_{i}) ] \\
& + \mathbb{P}[ \boldsymbol{d}_{i} \in \mathbb{B}(\text{DB}(f), \epsilon) , \pi_{f}(\boldsymbol{q}, \boldsymbol{d}_{i}) > \pi_{y}(\boldsymbol{q}, \boldsymbol{d}_{i}) ]. \\
\end{split}
\end{equation}
According to Eq.~\ref{tigeness gap} and Eq.~\ref{last bdy}, we can derive the gap $\mathcal{G}$ between the boundary ranking error and the upper bound of boundary ranking error.
Each term in the two equations includes a joint probability, which consists of:
\begin{enumerate*}[label=(\roman*)]
    \item the probability that document $\boldsymbol{d}_{i}$ within the neighborhood of the ranking decision boundary $\mathbb{B}(\text{DB}(f), \epsilon)$, and 
    \item the probability that the predicted ranking $\pi_f$ of ranking models has a specific positional relationship with the neighborhood ranking $\pi_n$ or ground truth ranking $\pi_y$. 
\end{enumerate*}

By considering the impact of the second part alone, we can reduce the joint probability into a marginal probability. 
Further, this allows us to bound the gap $\mathcal{G}$ as follows: 
\begin{equation}
\mbox{}\hspace*{-1mm}
\begin{split}
\mathcal{G} & \left(\mathcal{R}^*_{\mathrm{bdy}}(f), \mathcal{R}_{\mathrm{bdy}}(f) \right) \\
& = \mathcal{R}^*_{\mathrm{bdy}}(f) - \mathcal{R}_{\mathrm{bdy}}(f) \\
& = \mathbb{P}[ \boldsymbol{d}_{i} \in \mathbb{B}(\text{DB}(f), \epsilon) , \pi_{f}(\boldsymbol{q}, \boldsymbol{d}_{i}) < \pi_{n}(\boldsymbol{q}, \boldsymbol{d}_{i}) ] \\
& + \mathbb{P}[ \boldsymbol{d}_{i} \in \mathbb{B}(\text{DB}(f), \epsilon) , \pi_{f}(\boldsymbol{q}, \boldsymbol{d}_{i}) = \pi_{n}(\boldsymbol{q}, \boldsymbol{d}_{i}) ] \\
& + \mathbb{P}[ \boldsymbol{d}_{i} \in \mathbb{B}(\text{DB}(f), \epsilon) , \pi_{f}(\boldsymbol{q}, \boldsymbol{d}_{i}) > \pi_{y}(\boldsymbol{q}, \boldsymbol{d}_{i}) ] \\
& \leq \mathbb{P}[ \boldsymbol{d}_{i} \in \mathcal{D} , \pi_{f}(\boldsymbol{q}, \boldsymbol{d}_{i}) < \pi_{n}(\boldsymbol{q}, \boldsymbol{d}_{i}) ]  \\
& + \mathbb{P}[ \boldsymbol{d}_{i} \in \mathcal{D} , \pi_{f}(\boldsymbol{q}, \boldsymbol{d}_{i}) = \pi_{n}(\boldsymbol{q}, \boldsymbol{d}_{i}) ]  \\
& + \mathbb{P}[ \boldsymbol{d}_{i} \in \mathcal{D} , \pi_{f}(\boldsymbol{q}, \boldsymbol{d}_{i}) > \pi_{y}(\boldsymbol{q}, \boldsymbol{d}_{i}) ]  \\
& = \mathbb{E}_{\boldsymbol{d}_{i} \sim \mathcal{D}} \ \mathbb{I}\left\{ \pi_{f}(\boldsymbol{q}, \boldsymbol{d}_{i}) \  \neq \ \pi_{y}(\boldsymbol{q}, \boldsymbol{d}_{i}) \right\}. \\
\end{split}
\end{equation}
As a result, we deduce that the upper limit of the gap is the sum of three independent probabilities: for document $\boldsymbol{d}_{i}$,
\begin{enumerate*}[label=(\roman*)]
    \item the ranking position predicted by the ranking model is higher than the neighborhood ranking;
    \item the ranking position predicted by the ranking model is equal to the neighborhood ranking;
    \item the ranking position predicted by the ranking model is lower than the ground-truth ranking.
\end{enumerate*}

Therefore, the upper limit of the gap $\mathcal{G}$ is equivalent to the expectation that the ranking position predicted by the ranking model is not equal to the ground-truth ranking position. 
For this gap, its upper limit (note as $\mathcal{G}^*$) can be guaranteed due to the expectation being closely related to the natural ranking loss of a ranking model given by Eq. 9.

In summary, we have demonstrated that the gap $\mathcal{G}$ between the boundary ranking error $\mathcal{R}_{\mathrm{bdy}}(f)$ and its upper bound $\mathcal{R}^*_{\mathrm{bdy}}(f)$ is bounded, reflecting the tightness of the upper bound we provided. 
Moreover, $\mathcal{G}^*$, the upper limit of gap $\mathcal{G}$, is consistent with the natural ranking error, suggesting that this gap can potentially be optimized by the natural ranking loss.
In the following, we will present a detailed analysis to prove that the gap $\mathcal{G}$ is indeed optimizable under the natural ranking loss.

\heading{The natural optimizability of the gap} 
In this work, the natural ranking error is optimized through the natural ranking loss embedded within the proposed regularized surrogate loss as defined in Eq. 8. 
Our goal is to show that the upper limit of the ga $\mathcal{G}$p, denoted as $\mathcal{G}^*$, can be optimized through the natural ranking loss. 
This is achieved by demonstrating that the natural ranking error can be effectively optimized by the natural ranking loss.

Indeed, \citet{calauzenes2012non} have shown the effectiveness of natural surrogate losses, like the hinge loss, in optimizing classification errors. 
This insight was further extended by \citet{bartlett2006convexity}, who showcased that the pairwise loss function, acting as a natural ranking loss, optimizes the natural ranking error towards its infimum in conjunction with itself. 
In this paper, the pairwise loss we use for the natural ranking loss is also included. 
Hence, in the scenario of this paper, we have
\begin{equation}
\label{infimum}
\mbox{}\hspace*{-1mm}
\begin{split}
\mathcal{L}_{\mathrm{nat}}\left(\mathbb{D}, f\right) \rightarrow & \inf _{f} \mathcal{L}_{\mathrm{nat}}(\mathbb{D}, f) \\
& \Rightarrow \mathcal{R}_{\mathrm{nat}}\left(\mathbb{D}, f\right) \rightarrow \inf _{f} \mathcal{R}_{\mathrm{nat}}(\mathbb{D}, f),
\end{split}
\end{equation}
where the infima are taken over all data from the ranking dataset with the distribution $\mathbb{D}$.

According to Eq.~\ref{infimum}, the natural ranking error can be effectively optimized by our natural ranking loss.
Since $\mathcal{G}^*$, the upper limit of gap $\mathcal{G}$, is consistent with the natural ranking error.
It can also be naturally optimizable by the natural ranking loss.

\heading{The tightness of the upper bound of the boundary ranking error} 
Assume that the natural ranking error of the ranking model $f$ on the set of document candidates $\mathcal{D}$ is represented as, 
\begin{equation}
\label{eta}
\mbox{}\hspace*{-1mm}
\begin{split}
\mathbb{E}_{\boldsymbol{d}_{i} \sim \mathcal{D}} \ \mathbb{I}\left\{ \pi_{f}(\boldsymbol{q}, \boldsymbol{d}_{i}) \  \neq \ \pi_{y}(\boldsymbol{q}, \boldsymbol{d}_{i}) \right\} = \eta, 
\end{split}
\end{equation}
where $\eta \geq 0$ is the expectation of the natural ranking error associated with the ranking model $f$.
In this way, the upper limit of gap $\mathcal{G}$ can be determined as, 
\begin{equation}
\mbox{}\hspace*{-1mm}
\begin{split}
\mathcal{G} & \left(\mathcal{R}^*_{\mathrm{bdy}}(f), \mathcal{R}_{\mathrm{bdy}}(f) \right) \\
& \leq \mathcal{G}^* \left(\mathcal{R}^*_{\mathrm{bdy}}(f), \mathcal{R}_{\mathrm{bdy}}(f) \right) \\
& = \eta.\\
\end{split}
\end{equation}
Finally, according to Eq.~\ref{eta} and Eq.~\ref{infimum}, the upper limit of the gap $\mathcal{G}$ between the boundary ranking error $\mathcal{R}_{\mathrm{bdy}}(f)$ and its upper bound $\mathcal{R}^*_{\mathrm{bdy}}(f)$ is a deterministic value.
And it is confident that the deterministic value will be continuously reduced during training with the natural ranking loss.

Hence, given the presence of the deterministic value, optimizing the upper bound of the boundary ranking error $\mathcal{R}^*_{\mathrm{bdy}}(f)$ leads to effective optimization of the boundary ranking error $\mathcal{R}_{\mathrm{bdy}}(f)$ as well. 
Consequently, the upper bound of the boundary ranking error in Theorem 1 proves to be sufficiently tight.

\end{document}